\documentclass[runningheads]{llncs}
\usepackage{graphicx}
\usepackage{xcolor}
\usepackage{subfigure}
\usepackage{hyperref}
\usepackage{censor}
\usepackage{tokcycle}

\usepackage{moresize}
\usepackage{listings}
\AtBeginDocument{ 
  \counterwithout{lstlisting}{chapter}
}

\newcounter{censor}
\newcommand{\blind}[2]{{\ifthenelse{\value{censor}=1}{#2}{#1}}}
\newcommand{\alttext}[1]{{\colorbox{black}{\textcolor{white}{#1}}}}

\begin{document}
\blind{\title{Tool interoperability for model-based systems engineering\thanks{Research leading to these results has received funding from the EU ECSEL Joint Undertaking under grant agreement n\textsuperscript{o} 826452 (project Arrowhead Tools) and from the partners national programs/funding authorities.}}}{\title{Tool interoperability for model-based systems engineering\thanks{\xblackout{Research leading to these results has received funding from the EU ECSEL Joint Undertaking under grant agreement no 826452 (project Arrowhead Tools) and from the partners national programs/funding authorities.}}}}
%
%

\blind{\author{Sander~Thuijsman\inst{1} \and G\"{o}khan~Kahraman\inst{1} \and Alireza~Mohamadkhani\inst{1} \and Ferry~Timmers\inst{1} \and Loek~Cleophas\inst{1} \and Marc~Geilen\inst{1} \and Jan~Friso~Groote\inst{1} \and Michel~Reniers\inst{1} \and Ramon~Schiffelers\inst{1,2} \and Jeroen~Voeten\inst{1}}}{\author{Authors removed for double-blinded review\\[2\baselineskip]Blinded text appears as \censor{xxx} or \alttext{alternative text}\\[2\baselineskip]}}


%
\authorrunning{S.B. Thuijsman et al.}
%
\blind{\institute{Eindhoven University of Technology, Eindhoven, the Netherlands \and
ASML, Veldhoven, the Netherlands }}{\institute{ }}
\maketitle              
%

\begin{abstract}
Supervisory control design of cyber-physical systems has ma\-ny challenges. 
Model-based systems engineering can address these, with solutions originating from various disciplines. 
We discuss several tools, each state-of-the-art in its own discipline, offering functionality such as specification, synthesis, and verification.
Integrating such mono-discipli\-nary tools in a multi-disciplinary workflow is a major challenge. 
We present Analytics as a Service, built on the Arrowhead framework, to connect these tools and make them interoperable. 
A seamless integration of the tools has been established through a service-oriented architecture: The engineer can easily access the functionality of the tools from a single interface, as translation steps between equivalent models for the respective tools are automated.
\end{abstract}

\setcounter{footnote}{0}


%
%
%

\section{Introduction}
A Cyber-Physical System (CPS) forms a tight integration of cyber (computational) and physical components \cite{Lee2008}.
The physical system is monitored and controlled by (networks of) embedded computers.
Usually this occurs with a feedback loop: the computations affect the physical process and vice versa.
Examples of CPSs are automobiles, medical devices, and manufacturing systems.
In this work we consider the supervisory control design of manufacturing systems.
Supervisory control refers to the high-level (coordinated) monitoring and actuation of the system. 
In a manufacturing system, the supervisory controller regulates the manufacturing processes and the movement of products through the system.

The design of supervisory controllers may be performed by applying Model-Based Systems Engineering (MBSE), in which models, rather than documents, are the primary means of information exchange, and engineering processes are applied to these models directly \cite{Lee2008,Ramos2012}.
MBSE (of supervisory control) contains many different disciplines, among which specification, variation management, controller synthesis, optimization, formal verification, and implementation.
These disciplines each use a specific set of methods, tools, and technologies that are loosely coupled both on a syntactic and semantic level. 
This has a major impact on engineering efficiency: it hampers verification sufficiently early in the development process, especially concerning system-wide aspects such as throughput and collision avoidance. 
The simultaneous use and the integration of heterogeneous models and tools to capture system-wide properties reliably and with firm guarantees is an open issue \cite{Engell2015}.
To significantly improve engineering efficiency and enable rapid deployment of (new) systems and system features, seamless syntactic and semantic interoperability between engineering tools needs to be established \cite{Seshia2017}.


In this work we address how to create a multi-disciplinary workflow that has seamless integration of mono-disciplinary MBSE technologies.
We show how interoperability of MBSE tools can be achieved through \textit{Analytics as a Service (AaaS)}.
In AaaS, analytics functionality can be accessed over web-delivered technologies (i.e., the cloud).
Generally, AaaS is applied in the context of big data \cite{Sun2012,Demirkan2013,Assunccao2015,Skourletopoulos2016}.
In our work, however, we apply the concept of AaaS to MBSE:
key functionalities from various MBSE tools are offered as separate services on a network and are made interoperable by automatically translating models to equivalent ones that are necessary for the required service.
Furthermore, new functionality and tools can be added to the process in a modular manner.
The network is set up using the Arrowhead Framework, which is an service-oriented architecture for tool interoperability \cite{Varga2017,Venanzi2020}. 

In this paper we showcase and demonstrate the functionality of a number of state-of-the-art MBSE tools, and focus on the integration of these tools through AaaS to enable seamless synergistic multi-disciplinary MBSE.

%
%

To make the context more tangible, we first discuss a use case in Section \ref{sec:usecase}.
Examples from this use case are used throughout the paper.
In Section \ref{sec:challenges}, we discuss some challenges that emerge during the design of the supervisory control of CPSs.
Some state-of-the-art MBSE tools and technologies that address these challenges are discussed in Section \ref{sec:tools}.
These tools are made interoperable through AaaS, and integration of their functionality into a toolchain is discussed in Section \ref{sec:toolchain}.
Conclusions are provided in Section \ref{sec:conclusion}.



\section{Use case: xCPS manufacturing system}
\label{sec:usecase}
\begin{figure}[t]
\centering
\subfigure[xCPS system schematic overview.]{
\includegraphics[width=0.46\columnwidth]{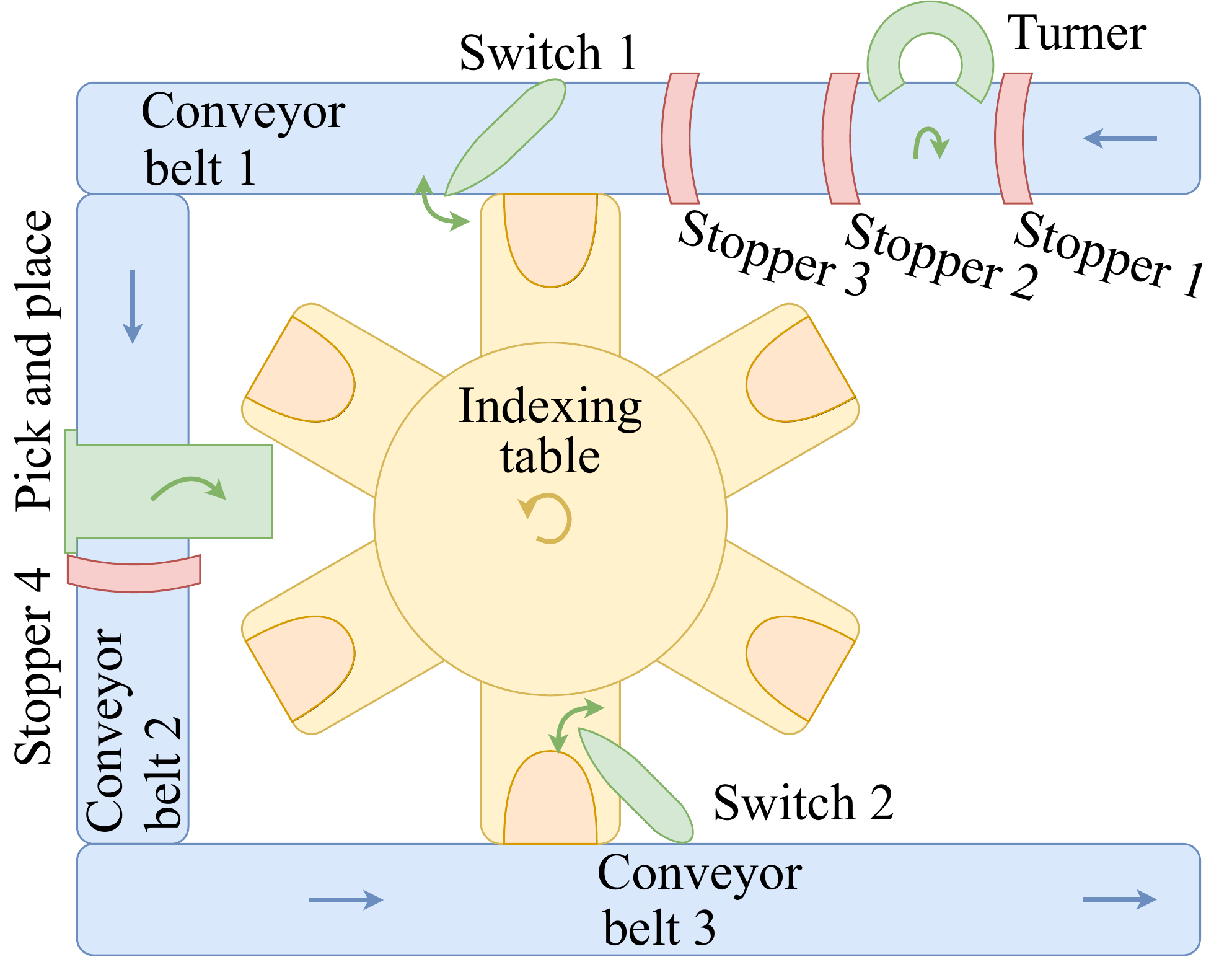}
\label{fig:overview}
}
\subfigure[xCPS system realization.]{
\includegraphics[width=0.46\columnwidth]{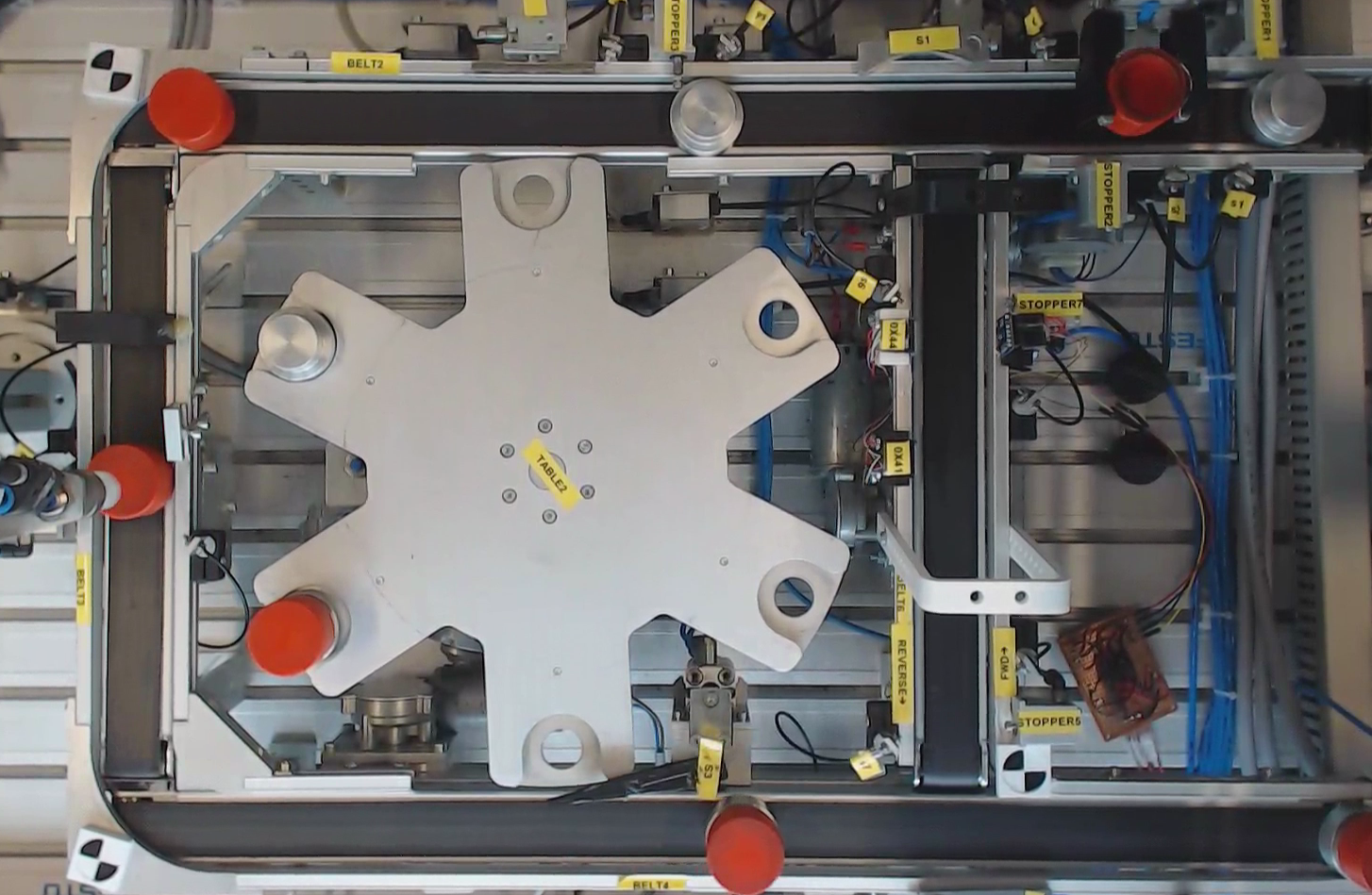}
\label{fig:realization}
}
\caption{xCPS system layout.}
\label{fig:layout}
\end{figure}

In this section we discuss a demonstrative use case from which we use examples throughout the paper.
Our method is demonstrated on the eXplore Cyber-Physical Systems (xCPS) manufacturing system.
It is a platform of industrial complexity for research and education on CPSs.
The xCPS system is elaborately discussed in \cite{Adyanthaya2017} and \cite{Basten2020}.
For simplicity, we only consider a part of the system, which is formed by the collection of components mentioned in Fig. \ref{fig:overview}.
The realization of this (sub)system is displayed in Fig. \ref{fig:realization}.

The xCPS system receives tops (red pieces in Fig. \ref{fig:realization}) and bottoms (silver pieces) on the right side of conveyor belt 1 in an alternating manner.
When a conveyor belt is moving, pieces can be held at a place by a stopper.
Pieces that are upside down can be turned with their right side up at the turner station.
Switch 1 can push pieces onto the indexing table, or allow pieces to keep running on conveyor belt 1.
The indexing table can hold up to six pieces (one on each arm), and turns counterclockwise.
Pieces that are not pushed on the indexing table transition to conveyor belt 2 where they reach the pick and place station.
To assemble a product, the pick and place robot picks up a top from conveyor belt 2, and places it onto a bottom on the indexing table.
When the indexing table turns after assembly, it brings the product to switch 2. 
Switch 2 can then push the assembled piece from the indexing table onto conveyor belt 3.
Finally, the assembled pieces leave the system at the right side of conveyor belt 3.

\section{Challenges}
\label{sec:challenges}
When controlling systems such as the xCPS system, several challenges may arise.
We mention some examples:
\begin{itemize}
\item How to manage variability in this system when there is for instance a system with, and a system without a turner station?
\item How to guarantee safe system operation, for instance, ensure a piece is never pushed to the indexing table when the spot on the table is already occupied by another piece?
\item How to control this system optimally to have the guaranteed highest throughput possible?
\item How to guarantee progress in the system, such that for every pair of top and bottom that enters the system, an assembled product will eventually leave the system?
\item After designing a controller for which the above guarantees can be made, how to deploy it on the system such that the guarantees are still made?
\end{itemize}

\begin{figure}[b!]
\centering
\includegraphics[width=1.00\columnwidth]{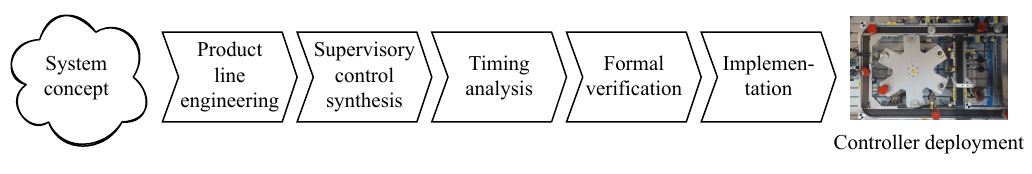}
\vspace*{-2\baselineskip}
\caption{Overview of the demonstrative workflow.}
\label{fig:workflow_simplified}
\vspace*{0.5\baselineskip}
\subfigure[In state-of-practice, manual processes are required to use functionality from different tools.\label{fig:AaaS1}]{
\centering
\includegraphics[width=0.475\columnwidth]{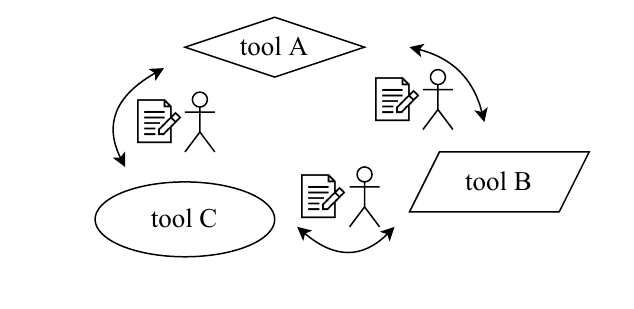}
}
\hfill
\subfigure[By applying AaaS, tools are made interoperable resulting in readily available functionality from each tool.\label{fig:AaaS2}]{
\centering
\includegraphics[width=0.475\columnwidth]{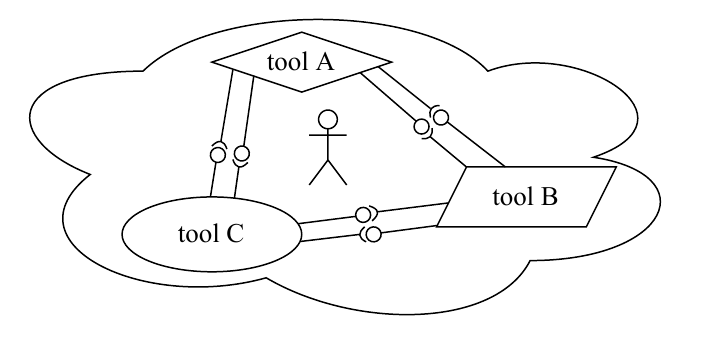}
}
\caption{From conventional MBSE (a) to AaaS (b).}
\end{figure}

Generally, (theoretical) solutions have been found for such challenges.
These solutions originate from various disciplines.
If we consider the challenges mentioned above, in respective order some relevant disciplines 
are: \textit{product line engineering} \cite{Pohl2005}, \textit{supervisory control synthesis} \cite{Cassandras2008}, 
\textit{timing analysis} \cite{Cohen1989}, \textit{formal verification} \cite{Baier2008}, and \textit{implementation} \cite{Dietrich2002}. 
Even though the challenges can be separately addressed, it is still a single system that is being engineered.
Therefore, a workflow is required that encompasses multiple disciplines.
A workflow in which the above challenges can be addressed is shown in Fig. \ref{fig:workflow_simplified}.

%

We observe that for each step in the workflow specialized functionality is required, which is present in distinct tools that each use their own syntax and semantics.
This makes it challenging to (sequentially) apply each step in a unified engineering process (Fig. \ref{fig:AaaS1}).
This brings us to the following major challenge that we address in this paper: 
\begin{itemize}
\item How to create a multi-disciplinary workflow that has seamless integration of mono-disciplinary MBSE technologies?
\end{itemize}
In this paper we present AaaS as a solution to tackle this challenge.
In this solution multiple tools are used, each specialized in their own discipline and functionality.
They are integrated (i.e., made interoperable) over a service-oriented architecture.
In this way, their functionality is offered through services, and the engineer can access them from a single interface (Fig. \ref{fig:AaaS2}).
When other tools or services are required, they may be offered as additional services for integration into the workflow.



\section{State-of-the-art tools and technologies}
\label{sec:tools}
In the following, we discuss several state-of-the-art tools that each exist in separate disciplines of MBSE.
The tools are LSAT, PLE tool, CIF, SDF3, mCRL2, Activity Execution Engine, and model translation tool.
These tools are just a selection of tools that could be applied in an MBSE process.
The mentioned tools are applied in \blind{the Arrowhead Tools}{\xblackout{the Arrowhead Tools}} project and are made Arrowhead framework compatible as discussed in Section \ref{sec:toolchain}.



\subsection{LSAT}
\label{sec:LSAT}
Logistics Specification and Analysis Tool (LSAT) has been developed in a collaboration between \blind{ESI (part of TNO, applied research center for high-tech systems design and engineering), ASML (manufacturer of lithography machines), and Eindhoven University of Technology (TU/e)}{\alttext{an applied research center, a manufacturing company, and a } \alttext{ university}}
\cite{Sanden2021}
\footnote{\blind{\url{https://lsat.esi.nl}}{\alttext{url 1}} , \url{https://projects.eclipse.org/projects/technology.lsat}}.
LSAT is being open sourced as an Eclipse project.
It is used for system specification through activity models \cite{Sanden2016,Thuijsman2020LSAT}.
A system is represented by a number of resources, where each resource consists of a number of peripherals.
Each peripheral can execute a set of actions, to which a timing is prescribed. 
An activity describes a cohesive piece of behavior in the system, and is modeled as a directed acyclic graph that captures the dependencies between actions performed on peripherals of the resources that it uses. 
Additionally, LSAT has visualization techniques such as: movement trajectory plots for moving peripherals, graphical editing of activities, and Gantt charts to represent the timing of a sequence of activities.

In Listing \ref{lst:LSATmachine}, an LSAT definition for three peripherals of the xCPS system is given: \texttt{gripper}, \texttt{turner}, and \texttt{zMotor}.
For these peripherals, either actions are defined which they can perform, or axes are defined along which they can move.
The peripherals are instantiated in resource \texttt{Turner}.
The instantiation of the \texttt{zMotor} is parameterized with positions it can move to (\texttt{Above\_Belt}, \texttt{At\_Belt}), and which movements are possible between these positions (in this case in both directions between both positions), with a speed profile.

\begin{lstlisting}[frame=single,caption=LSAT machine specification.\label{lst:LSATmachine}]
PeripheralType gripper {
	Actions {
		grab
		ungrab
	}
}

PeripheralType turner {
	Actions {
		flip_left
		flip_right
	}
}

PeripheralType zMotor {
  SetPoints {
    Z [m]
  }
  Axes { 
    Z [m] moves Z
  }
}

Resource Turner {
	turner : turner 
	gripper : gripper
	zMotor : zMotor {
		AxisPositions {  
      Z (Above,At) 
		}
		SymbolicPositions {
			Above_Belt (Z.Above)
			At_Belt (Z.At)
		}
		Profiles (normal)
		Paths {
			Above_Belt <-> At_Belt profile normal 
		}
	}
}
\end{lstlisting}

In Listing \ref{lst:LSATsetting}, part of the LSAT setting specification of the xCPS system is shown.
The timings of the actions of the \texttt{gripper} and \texttt{turner} peripherals in the \texttt{Turner} resource are defined.
These timings are deterministic, but LSAT also supports specification of probability distributions for timing.
For the \texttt{zMotor} of the \texttt{Turner} a speed profile \texttt{normal} is defined by specifying (maximal) velocity, acceleration, and jerk. 
Coordinates are defined for the symbolic positions of the peripheral.
While specifying activities, the user can define actions that move the peripheral between these physical positions with a specific speed profile. 
Then, LSAT computes the timing of that action.

\begin{lstlisting}[frame=single,caption=LSAT setting specification.\label{lst:LSATsetting}]
Turner.gripper {
	Timings {
		grab = 0.05
		ungrab = 0.04
	}
}

Turner.turner {
	Timings {
		flip_left = 0.35
		flip_right = 0.35
	}
}

Turner.zMotor {
	Axis Z {
		Profiles {
			normal (V = 5, A = 10, J = 10)
		}
	Positions {
		Above = 0
		At= 0.12
    }
  }
}
\end{lstlisting}

In Listing \ref{lst:LSATactivity}, an LSAT specification for the activity \texttt{TurnerTurnTop} is given.
This is done by first giving arbitrary names for the actions that are used in the activity, and then specifying the directed acyclic graph of the activity which defines its action flow.
An arrow (\texttt{->}) between two actions denotes that the succeeding action always starts after the preceding action has completed.
Synchronization points are marked by a vertical bar: \texttt{|s1} and \texttt{|s2} are used in Listing \ref{lst:LSATactivity} to denote multiple incoming or outgoing dependencies for an action.
E.g., after \texttt{Down}, both actions \texttt{Release} and \texttt{Up2} are allowed to take place (in any order/at the same time).
In the activity framework, each resource needs to be claimed by the activity before it can perform actions, and all claimed resources need to be released after performing the actions \cite{Sanden2016}.
In this way, activities can be deployed in a pipelined manner.
I.e., when a resource is free, an activity can claim it and perform its actions on that resource, regardless of what previous activities might still be performing (on other resources). 

\begin{lstlisting}[frame=single,caption=LSAT activity specification.\label{lst:LSATactivity}]
activity TurnerTurnTop {
	prerequisites{
		Turner.zMotor at At_Belt
	}
	actions {  
		CT1     : claim Turner
		RT1     : release Turner
		CS1     : claim Stopper1
		RS1     : release Stopper1  
		CS2     : claim Stopper2
		RS2     : release Stopper2  
		Left    : Turner.turner.flip_left 
		Right   : Turner.turner.flip_right
		Up      : move Turner.zMotor to Above_Belt with speed profile normal 
		Up2     : move Turner.zMotor to Above_Belt with speed profile normal
		Down    : move Turner.zMotor to At_Belt with speed profile normal
		Grab    : Turner.gripper.grab
		Release : Turner.gripper.ungrab  
	}
	action flow {
    CS2->CS1->CT1->Grab->Up->Left->Down->|s1->Release->|s2->Right->RT1->RS2->RS1
                                         |s1->Up2    ->|s2
	} 
}
\end{lstlisting}

\subsection{PLE tool}
The Product Line Engineering (PLE) tool, developed at \blind{TU/e}{\alttext{a university}}, is used to manage the variability of a system, and to automatically validate and derive product instances within a product line 
\cite{Kahraman2021}.
Using the PLE tool, a product line is defined which consists of a feature model, a base LSAT model, delta modules, and a mapping model. 

\begin{figure}[ht]
\centering
\includegraphics[width=0.90\columnwidth]{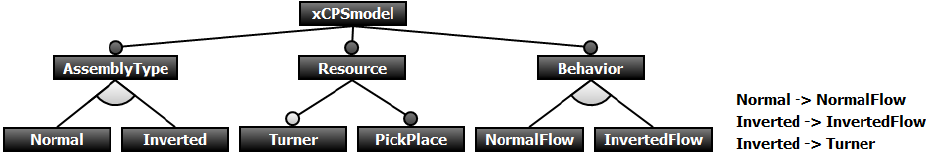}
\caption{xCPS feature model}
\label{fig:feature_model}
\end{figure}

The feature model expresses the commonality and variability of the product line \cite{Benavides2010}.
Fig. \ref{fig:feature_model} shows the feature model of the xCPS product family that represents the variability in the resources, behavior, and assembly type of the system.
Constraints expressed as propositional formulas are used to define dependencies between features. 
A configuration is valid if the combination of features is allowed by the feature model.


A base LSAT model, such as described in Section \ref{sec:LSAT}, is input to the PLE tool and serves as source for deriving new product instances. 
Delta modules are used to define modifications that can be made to the base model.
In Listing \ref{lst:PLEdelta} a delta module of the LSAT machine specification is provided for the absence of the turner station.
Note that for example the gripper peripheral is not removed, since it is used in more resources next to the turner.
\begin{lstlisting}[frame=single,caption=PLE Tool delta module.\label{lst:PLEdelta}]
delta "machineDelta"
	dialect <http://www.esi.lsat.nl/machine>
	modifies <../model/xCPS.machine>
{	removeResourceFromResourcesOfMachine(<Turner>, <xCPS>);
	removePeripheralFromPeripheralTypesOfMachine(<turner>, <xCPS>);
	... }
\end{lstlisting}
The PLE tools uses mapping models to define what delta modules need to be applied when a particular configuration is selected.
In Listing \ref{lst:PLEmapping} a mapping model is provided, where \texttt{!Turner} states that the following deltas should only be applied if the Turner is not selected.
\begin{lstlisting}[frame=single,caption=PLE Tool mapping model.\label{lst:PLEmapping}]
!Turner:
  <deltas/machineDelta.decore>,
  <deltas/activityDelta.decore>,
  ...
\end{lstlisting}
A particular instance of the xCPS product line is defined in Listing \ref{lst:PLEconfiguration}.
Note that next to the absence or presence of components, the PLE tool can also be used to, e.g., instantiate settings (for the example, defined in \texttt{FastMovement}) or assembly procedures.
Using the defined product line and the configuration, the PLE tool derives LSAT variant models.
These derived models can then be used to perform further analysis.
\begin{lstlisting}[frame=single,caption=PLE Tool configuration.\label{lst:PLEconfiguration}]
configuration <xCPS.defeaturemodel> {
	"Resource",
	"PickPlace",
	"Turner",
	"Behavior",
	"FastMovement",
}
\end{lstlisting}

\subsection{CIF} 
\label{sec:CIF}
CIF is an automata-based tool and language, and is used to specify system behavior, formulate behavioral requirements, and perform supervisory controller synthesis \cite{Cassandras2008} to obtain a correct-by-construction supervisory controller that adheres to the requirements
\cite{vanBeek2014}.
CIF is part of the Eclipse Supervisory Control Engineering Toolkit (Eclipse ESCET™) \cite{ESCET}\footnote{\url{https://www.eclipse.org/escet/} , `Eclipse', ‘Eclipse ESCET’ and ‘ESCET’ are trademarks of Eclipse Foundation, Inc.}, that has become an Eclipse open source project since 2020.
This project builds upon research and tool development at \blind{TU/e, as well as collaboration with industry including ASML, and Rijkswaterstaat (part of the Dutch ministry of infrastructure and water management).}{\alttext{a university}, as well as collaboration with industry including \alttext{a manufacturing company}, and \alttext{an infrastructural company}.}



In Listing \ref{lst:CIFrequirement} a requirement specified in CIF is shown.
The requirement automaton specifies allowed behavior for the turner station.
Activities \texttt{TopGoTo}\-\texttt{Turner} and \texttt{BottomGoToTurner} can only occur when there is no piece present at the turner station.
A bottom can immediately pass through the turner station.
It is assumed that every top needs to be inverted.
First, the \texttt{TurnerGoDown} activity is executed.
This activity can be successful or fail.
When it is successful, the turner will turn the top, and the top can continue after the turner.
When the activity fails, the turner has to retry until it is successful.

\begin{lstlisting}[frame=single,caption=CIF requirement turner.\label{lst:CIFrequirement}]
requirement automaton TurnerFlow:
	location NoPiece:
		initial; marked;
		edge TopGoToTurner goto TurnTop;
		edge BottomGoToTurner goto Bottom;
	location TurnTop:
		edge TurnerGoDown goto TurningTop;
	location TurningTop:
		edge TURNERDOWNSUCC_TurnerTurnTop goto TurnedTop;
		edge TURNERDOWNFAIL_RetryTurnerGoDown;
	location TurnedTop:
		edge TopGoAfterTurner goto NoPiece; 
	location Bottom:
		edge BottomGoAfterTurner goto NoPiece;
end
\end{lstlisting}
%
%

CIF can be used to perform supervisory controller synthesis. 
A supervisory controller is generated, that restricts the behavior of the system such that the requirements are always satisfied.
Essentially, for each activity a predicate is computed that needs to hold for the activity to occur.
For example, Listing \ref{lst:CIFsupervisor} shows the additional guard that is generated for the activity \texttt{BottomGoAfterTurner}.
It makes sure that this activity can only occur when at the next station after the turner (\texttt{Sensor3}) there are no inverted tops present, of which the amount is stored in \texttt{Sensor3Location.nInvTops}, to avoid collisions.
Note that when \texttt{BottomGoAfterTurner} occurs, there can never be a non-inverted top or bottom at Sensor3, since it is assumed that tops and bottoms are input in alternating manner, cannot overtake, and every top is inverted at the turner station.

\begin{lstlisting}[frame=single,caption=CIF supervisor.\label{lst:CIFsupervisor}]
supervisor automaton sup:
  location:
    initial; marked;
    edge BottomGoAfterTurner when Sensor3Location.nInvTops = 0;
    ...
end
\end{lstlisting}

\subsection{SDF3} 
\label{sec:SDF3}
SDF3 is a toolset that has an extensive library of analysis and transformation algorithms for synchronous dataflow graphs, 
which are suitable for modeling both parallel and pipelined processing and cyclic dependencies \cite{Stuijk2006}. 
Additionally, it can be used to generate random synchronous dataflow graphs, if desirable with certain guaranteed properties.
SDF3 is developed at \blind{TU/e \footnote{\url{https://www.es.ele.tue.nl/sdf3/}}}{\alttext{a university} \footnote{\alttext{url}}}.
In this work we will use it for makespan optimization of activity models to find the optimal behavior that produces products as quickly as possible.


From an Input/Output (I/O) automaton \cite{lynch1988introduction} and a set of max-plus matrices that capture 
the timing and behavior of our system with event feedback \cite{mohamadkhani}, SDF3 makes an internal conversion to a max-plus automaton
and performs 
timing optimization using the methods of \cite{Gaubert1995} to find the optimal order of activities to be dispatched to generate the lowest possible makespan.
In Listing \ref{lst:SDF3MPA} an I/O automaton is shown. 
\texttt{loc1} is the initial location (denoted by \texttt{i}).
Transitions between the locations are defined, where in this case most input actions are empty (denoted by an empty string).
\texttt{loc4} is where the event outcome of \texttt{TurnerGoDown} is processed. 
If it succeeds, the top piece is turned, and if it fails, the turner retries to go down.

For the xCPS system model, the computed dispatching sequence is shown in Listing \ref{lst:SDF3result}.
This is the optimal sequence of activities to execute when the system is empty to produce one product as quickly as possible.
When there is no fail on the turner (or elsewhere), one product can be produced in 17.723 seconds.

\begin{figure}[t]
\centering
\includegraphics[width=1.00\columnwidth]{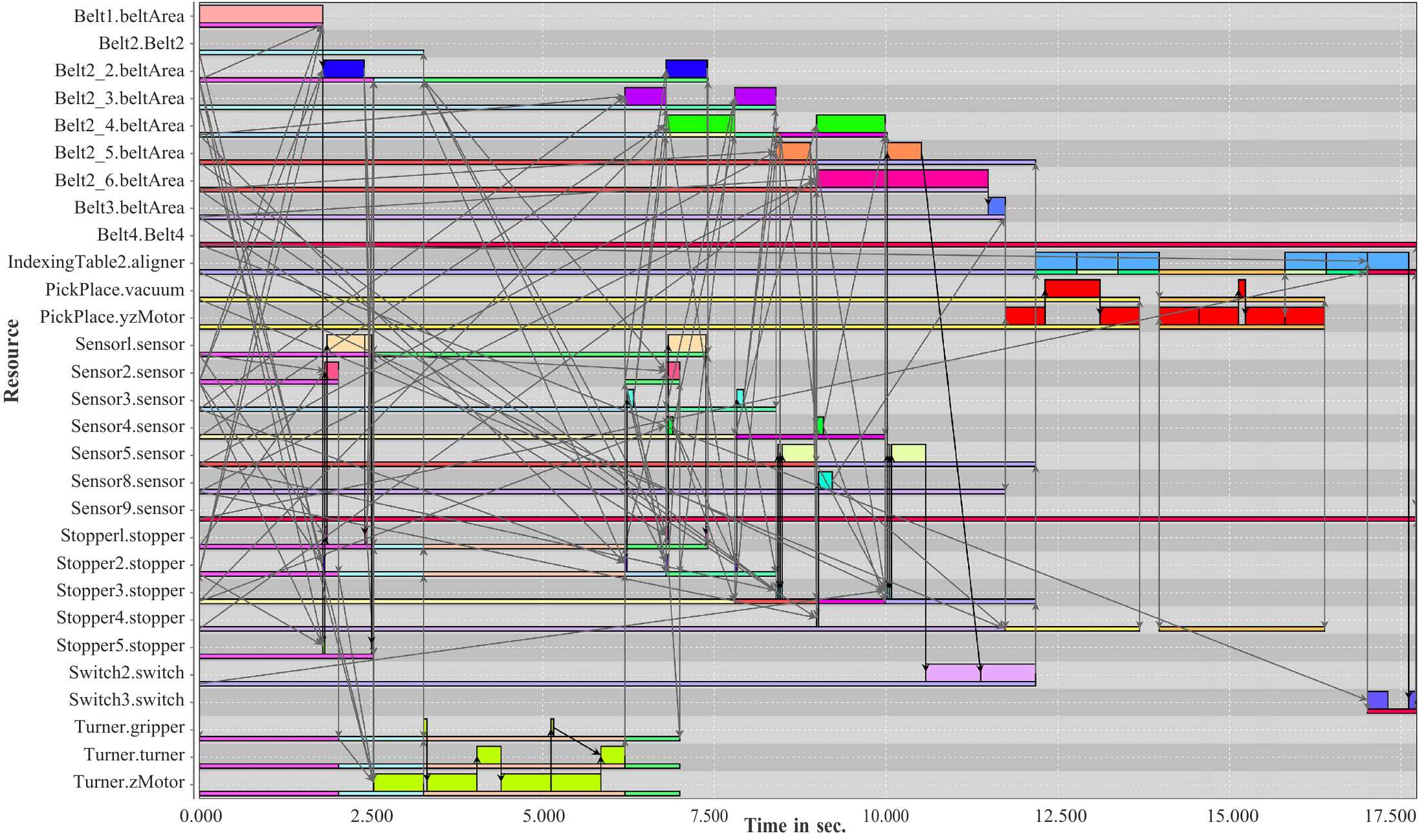}
\vspace*{-1.2\baselineskip}
\caption{Gantt chart optimal makespan one product.}
\vspace*{-0.2\baselineskip}
\label{fig:Gantt}
\end{figure}

\begin{lstlisting}[frame=single,caption=SDF3 I/O automaton.\label{lst:SDF3MPA}]
ioautomaton statespace {  
  loc1 i -,InputTopInverted-> loc2 
  loc2 -,TopGoToTurner-> loc3 
  loc3 -,TurnerGoDown-> loc4 
  loc3 -,InputBottom-> loc5  
  loc4 -TURNERDOWNSUCC,TurnerTurnTop->  loc6 
  loc4 -TURNERDOWNFAIL,RetryTurnerGoDown->  loc7
  loc4 -,InputBottom-> loc8
  ... }
\end{lstlisting}
\begin{lstlisting}[frame=single,caption=SDF3 makespan result.\label{lst:SDF3result}]
makespan: 17.723
sequence: InputTopInverted; TopGoToTurner; InputBottom; ; TurnerGoDown; TURNERDOWNSUCC,TurnerTurnTop; TopGoAfterTurner; BottomGoToTurner; TopGoToSensor4; ; BottomGoAfterTurner; TopGoToSwitch2; BottomGoToSensor4; BottomGofromSt3ToTable2; AlignTable2WithPickPlace; AlignTable2WithBelt; AlignTable2WithPickPlace; TopGoToPickPlace; TopPickUp; TopAssemble; AlignTable2WithBelt; AlignTable2WithPickPlace; ProdGoFromTable2ToBelt4
\end{lstlisting}
To study the sequence and evaluate potential bottlenecks, LSAT can generate a Gantt chart from the sequence, which is shown in Figure \ref{fig:Gantt}.
The Gantt chart shows which activities are occupying which resources at what time, what actions are executed, and the dependencies between actions on different peripherals.

\subsection{mCRL2}
mCRL2 is a toolset designed to reason about concurrent and distributed systems.
The toolset consists of its own language as well as more than sixty tools supporting visualization, simulation, minimization, and model checking \cite{Baier2008} of complex systems
\cite{Groote2014,Bunte2019}
\footnote{\url{https://www.mcrl2.org/}}.
mCRL2 is open source and developed at \blind{TU/e}{\alttext{a university}} in collaboration with \blind{University of Twente}{\alttext{another university}}. 
Given an mCRL2 specification and a property in the modal mu calculus, mCRL2 can apply model checking to guarantee the (non-)existence of particular behaviors in a system.

A portion of an mCRL2 model describing the behavior relevant to the turner station and the activity TurnerTurnTop is provided in Listing \ref{lst:mCRL2model}.
The model is derived from an automata translation of the LSAT model using \cite{Thuijsman2020LSAT}.

\begin{lstlisting}[frame=single,caption=mCRL2 model.\label{lst:mCRL2model}]
sort enum_LPE = struct enumlit_Grab | enumlit_Ungrab;
sort enum_LPE2 = struct enumlit_Flip_left | enumlit_Flip_right;
...
act value_Turner_gripper : enum_LPE;
act value_Turner_turner : enum_LPE2;
...
proc BehProc_M(Locvar_M : LocSort_M, Activity_TurnerTurnTop : enum_LPE4, Turner_gripper : enum_LPE, Turner_turner : enum_LPE2, Turner_zMotor : enum_LPE3) =
  value_Activity_TurnerTurnTop(Activity_TurnerTurnTop) . BehProc_M(Locvar_M, Activity_TurnerTurnTop, Turner_gripper, Turner_turner, Turner_zMotor) +
  value_Turner_gripper(Turner_gripper) . BehProc_M(Locvar_M, Activity_TurnerTurnTop, Turner_gripper, Turner_turner, Turner_zMotor) +
  ...
  ((Locvar_M == loc_M_L) && (Activity_TurnerTurnTop == enumlit_l0)) -> claim_Stopper2 . BehProc_M(Locvar_M, enumlit_l1, Turner_gripper, Turner_turner, Turner_zMotor) +
  ((Locvar_M == loc_M_L) && (Activity_TurnerTurnTop == enumlit_l1)) -> claim_Stopper1 . BehProc_M(Locvar_M, enumlit_l2, Turner_gripper, Turner_turner, Turner_zMotor) +
  ... ;
act claim_Stopper1, renamed_claim_Stopper1, claim_Stopper2, ...;
init BehProc_M(loc_M_L, enumlit_l0, enumlit_Ungrab, enumlit_Flip_left, enumlit_Above_belt);
\end{lstlisting}

In Listing \ref{lst:mCRL2requirement}, a modal mu calculus formula is provided (in mCRL2 syntax) that expresses that for every time the turner flips left, the turner must eventually flip right. 
mCRL2 can be used to verify whether this property holds for the model.
When a property that is checked does not hold, a counter-example is given to aid in solving the problem.
\begin{lstlisting}[frame=single,caption=mCRL2 requirement.\label{lst:mCRL2requirement}]
[true*.flip_left]mu X.([!flip_right]X && <true>true)
\end{lstlisting}

\subsection{Activity Execution Engine}
The Activity Execution Engine (AEE) is an automated execution method for the activity framework developed at \blind{TU/e}{\alttext{a university}}.
It receives an activity model with a supervisory controller in the form of an I/O automaton as input and executes it on the system.
The AEE is time-preserving, meaning it adheres to (LSAT) model prescribed timing of actions within well defined bounds.
The AEE guarantees determinate behavior of the system despite timing variations that may happen in execution.
The supervisory control is directly performed by the activity execution engine, i.e., the activity execution engine directly connects to the low-level resource controllers.

The execution engine provides a generic solution for executing the LSAT model on the machine. 
It consists of three layers: 
The supervisory control layer deals with the high-level execution concerning the order of activities and decisions based on event outcomes that it receives from the lower layers.
The AEE layer is concerned with execution of individual activities and sequencing them together. 
The action translation layer is responsible for translating action descriptions in LSAT specification to low-level function calls that execute the actions, and translates the sensor data back to communicate to the higher levels. 

The generic solution provides all the mechanisms needed to guarantee timing and behavior-preserving execution of the model.
The only part of the engine that the system designers and engineers would need to implement based on their specific product is the action translation layer which is a relatively small library.

\subsection{Model translation tool}
As becomes apparent from the above listings, each tool has its own unique syntax.
To make the tools interoperable, it needs to be possible to translate a model in one tool to an equivalent model in another tool.
In this way it is for example possible to generate a behaviorally equivalent mCRL2 model of an LSAT model and perform formal verification on that model, which is a function of only mCRL2. 
These translations are essential for the approach that we discuss next in Section \ref{sec:toolchain}.
It allows to automate the use of functionality of tool \emph{X} for models in (syntax of) tool \emph{Y}.
The model translation tool is a toolset that offers a range of translations between models that are equivalent with respect to the process or computation that will be performed on the generated model.
For models with complex semantics, performing translation validation \cite{Pnueli1998} may be desirable.

\section{Toolchain and interoperability}
\label{sec:toolchain}
The introduced tools each have their own specific functionality.
With their combined functionality, they can be used in a MBSE workflow for manufacturing systems.
We first discuss a workflow that uses all tools mentioned in Section \ref{sec:tools}, and then present how the tools are integrated in an interoperable toolchain.

\subsection{Example workflow}
\label{sec:workflow}
In Section \ref{sec:challenges}, we discussed a number of challenges that arise in MBSE of supervisory controllers.
An interdisciplinary workflow was introduced in Figure \ref{fig:workflow_simplified}.
In that workflow, each of the steps tackled a challenge in the engineering process.
All steps of the workflow can individually be performed by the tools mentioned in Section \ref{sec:tools}.
In this section we discuss integration of the tools into a toolchain, creating a MBSE process of supervisory control design for manufacturing systems that includes specification, variation management, supervisor synthesis, timing optimization, formal verification, and implementation.

We use the following workflow to show how the tools can be integrated together to form a toolchain:
\begin{enumerate}
\item A product line is specified using LSAT and the PLE tool.
\item Given a feature configuration, an LSAT product instance is derived with the PLE tool.
\item The LSAT product instance specification is converted to CIF.
\item Safety requirements are specified, and a maximally permissive supervisory controller automaton is synthesized using CIF.
\item The automaton supervisor is converted to an I/O automaton.
\item Using the I/O automaton and the timing information from the LSAT specification, SDF3 is used to perform makespan optimization to find the optimal dispatching sequence of activities to produce products as quickly as possible.
\item The LSAT specification and the optimal dispatching sequence are used to construct an mCRL2 model.
\item With mCRL2, progress properties are verified for the behavior that results from the obtained dispatching sequence.
\item The control strategy is deployed on the physical system using the AEE.
\end{enumerate}
The authors note that this is just a demonstrative workflow, the tools provide more services that could be used, or different tools can be used altogether.

\begin{figure}[tbhp]
\centering
\includegraphics[width=1.00\columnwidth]{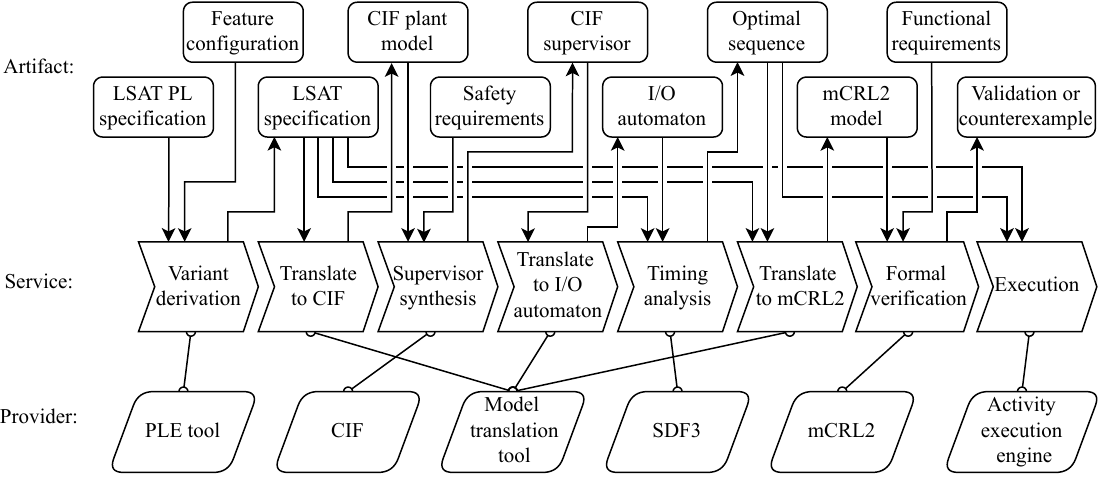}
\caption{Elaborated overview demonstrative workflow.}
\label{fig:workflow}
\end{figure}

A graphical representation of the workflow is shown in Figure \ref{fig:workflow}. 
Essentially, this is an elaboration of the workflow in Figure \ref{fig:workflow_simplified}.
At the top of the diagram are the artifacts.
Some artifacts are manually constructed (these do not have an incoming arrow), others are automatically generated during the process.
In the middle are services corresponding to processes that are performed during the workflow, with incoming and outgoing arrows linking them to the respective input and output artifacts. 
The processes are executed as a service provided by one of the tools shown in the bottom of the diagram.

Excerpts from the artifacts are presented throughout Section \ref{sec:tools} \footnote{Complete models of the xCPS system for the various tools are available here: \\ \url{https://github.com/sbthuijsman/FM_interoperability}}.

Through this workflow, the challenges mentioned in Section \ref{sec:challenges} are addressed.
Using the PLE tool, variability of the system is managed by specifying a product line with deltas between particular configurations.
In this way, we avoid (manually) creating models of each unique configuration.
Behavioral requirements are specified in a modular manner, and a minimally restrictive correct-by-construction supervisory controller is generated that adheres to these requirements by applying supervisory controller synthesis in CIF.
Time-optimal control is guaranteed through application of timing optimization using SDF3, which selects the time-optimal activity sequence from the supervisory controller.
Progress in the system can be guaranteed by specifying and verifying progress properties using mCRL2.
Finally, the designed controller is deployed on the system with the AEE, which guarantees execution that adheres to the models.
Even though these tools use their own semantics and syntax, their functionality can be applied sequentially on a single system through the use of model translations by using the model translation tool.

The authors note that even though behavioral requirements are already enforced by supervisor synthesis through CIF, formal verification in mCRL2 is still beneficial.
A progress property such as listed in Listing \ref{lst:mCRL2requirement} (eventually modality) can not be directly expressed as a requirement in CIF, and supervisory controller synthesis is in general not applicable to such requirements.
Additionally, in this case the CIF model only considers behavior on activity level, while the mCRL2 model includes action level behavior, which allows for more detailed inspection of the behavior.



\subsection{Toolchain integration over Arrowhead framework}
\label{sec:arrowhead}
The workflow described above spans from specification to realization.
In the steps along the way, several processes are performed using various tools.
A seamless integration of these tools is established through AaaS: their functionalities are provided as services on an Arrowhead local cloud.
The Arrowhead cloud is constructed using the Arrowhead framework, which is an interoperability service-oriented architecture \footnote{\url{https://arrowhead.eu/} , \url{https://github.com/eclipse-arrowhead} , the Arrowhead framework files used for the integration of the tools discussed in this paper can be found here: \url{https://github.com/sbthuijsman/FM_interoperability}}.
The framework is elaborately discussed in \cite{Varga2017,Venanzi2020}. 
Next to the MBSE tool services, the following three Arrowhead core services exist on the cloud:
\begin{enumerate}
\item Orchestration service to coordinate the connections between the consumer and provider services.
\item Authorization service to provide security such that services can only be accessed by authorized consumers.
\item Service registry system to keep track of all services within the network and ensure all systems can find each other.
\end{enumerate}
Because of the integration in the Arrowhead cloud, the workflow can be performed from a single interface and the functionality from each tool is readily accessible.
Furthermore, additional tools and services can be added in a modular manner.
As long as there is a tool or service that can solve the problem, and model-to-model translation is possible from some existing artifact to the required syntax of the concerning service, the toolchain can be extended to include the service in the manner as the discussed services.
Direct access to functionality of the MBSE tools enables rapid application of MBSE technologies from various disciplines, and allows seamless MBSE of the supervisory control of a CPS from start to finish.
Next to this integration, the AaaS framework provides more benefits, such as provisioning of a centralized (model) database or management of computational resources.

\section{Conclusion}
\label{sec:conclusion}
Many challenges can be addressed through MBSE of the supervisory control of CPSs.
Solutions originate from several disciplines.
Each discipline uses its own tools with its own semantics and syntax.
It is a major challenge to create a multi-disciplinary workflow that has seamless integration of mono-disciplinary MBSE technologies.

In this paper, several tools are discussed that are each state-of-the-art in their own discipline.
Even though the tools use their own semantics and syntax, equivalent models can be generated for each tool by model-to-model translations.
By applying AaaS, in this case using the Arrowhead framework, the tools are made interoperable.
The translation steps are automated, and the services from each tool are readily accessible.
A seamless integration of the tools is established: the engineer can easily access their functionality from a single interface.
Because of the modularity of the service-oriented architecture, the toolchain can without difficulty be extended to incorporate additional functionality as long as the required model-to-model translations can be established.

%
%
%
 \bibliographystyle{splncs04}
 \bibliography{references}

\end{document}